\documentclass[preprint,12pt,authoryear]{elsarticle}

\usepackage{amssymb}
\usepackage{color}
\usepackage{graphicx}
\usepackage{lineno}
\usepackage{amsmath}
\usepackage{subfig}
\usepackage{rotating}
\usepackage[round]{natbib} 

\begin{document}

\begin{frontmatter}

 \title{Impact of dispersal on the stability of metapopulations} 
 \author[]{Eric Tromeur\corref{cor1}} 
 \ead{eric.tromeur@ens-lyon.fr}

 \cortext[cor1]{\emph{Phone}: +33644272248}

\author{Lars Rudolf}
\author{Thilo Gross}

\address{University of Bristol, Merchant Venturers School of Engineering, Bristol, UK}

\begin{abstract}
Dispersal is a key ecological process, that enables local populations to form spatially extended systems called metapopulations.
In the present study, we investigate how dispersal affects the linear stability of a general single-species metapopulation model.
We discuss both the influence of local within-patch dynamics and the effects of various dispersal behaviors on stability.
We find that positive density-dependent dispersal and positive density-dependent settlement are destabilizing dispersal behaviors while negative density-dependent dispersal and negative density-dependent settlement are stabilizing. 
It is also shown that dispersal has a stabilizing impact on heterogeneous metapopulations that correlates positively with the number of patches and the connectance of metapopulation networks.
\end{abstract}


\end{frontmatter}


\section{Introduction}
\label{intro}

Many species occupy disconnected habitats that consist of individual patches linked by dispersal \citep{MacArthurWilson1967}.
Understanding the dynamics of such species requires describing them as a meta-population that is formed of populations in the respective patches \citep{Levins1969,Hanski1999}.
Untangling the influence of dispersal on the stability of such metapopulations is a major challenge. 

While the investigation of metapopulation dynamics classically relies on extinction-colonization models \citep{Hanski1998}, recent progress has been made by extending population-dynamical models to the metapopulation context.
Dispersal in metapopulations has been found to be both stabilizing and destabilizing, depending on the intensity of dispersal \citep{BriggsHoopes2004}. It is generally thought that weak dispersal stabilizes metapopulations by generating asynchronous dynamics between patches \citep{Taylor1990,Ruxton1994,BriggsHoopes2004}, while strong dispersal is expected to destabilize metapopulations by promoting greater synchrony between patches \citep{Hastings1993,Ruxton1994}. However, it has been shown that metapopulation synchrony and stability may also be positively correlated \citep{Abbott2011}. Further, the influence of dispersal on stability is crucially mediated by local dynamics and dispersal behaviors \citep{Amarasekare1998,Amarasekare2004}. 

Dispersal behaviors are life history traits that affect the fitness of individuals in heterogeneous landscapes \citep{Dieckmann1999}. Although dispersal has long been modelled as a linear, density-independent behavior \citep{BascompteSole1994}, it now appears that density-dependent dispersal is a widespread strategy \citep{BowlerBenton2005}, that can appear as a result of eco-evolutionary dynamics \citep{Travis1999}. Dispersal-related behaviors take various forms, in all steps of dispersal: emigration, inter-patch movement and immigration \citep{BowlerBenton2005}. For instance, emigration can be triggered by an overcrowded patch and immigration can be enhanced or inhibited by a high density of conspecifics. While adaptive, these behaviors may also affect the dynamics of metapopulations and bring local populations on the edge of extinction \citep{Dieckmann1999}.

Bascompte \& Sol\'{e} \citeyearpar{BascompteSole1994} found that increasing density-independent dispersal can destabilize metapopulations, whereas \citet{Hassell1995}, followed by \citet{Rohani1996} and Jang \& Mitra \citeyearpar{JangMitra2000}, concluded that it does not influence stability.
Further, it has been shown that under certain conditions, density-dependent dispersal can be destabilizing \citep{Ruxton1996,Silva2001,SilvaGiordani2006}.
By contrast, \citet{Ruxton1994} and Stone $\&$ Hart \citeyearpar{StoneHart1999} argued that a weak coupling between chaotic patches can stabilize metapopulations and \citet{Ruxton1997a} subsequently found that also costly dispersal has a stabilizing effect.
This result has been questioned by \citet{Kisdi2010}, who showed that costly dispersal can also have a destabilizing influence on metapopulation dynamics, using a specific growth function.

Dispersal behaviors relating to arrival and settlement of immigrants in new patches can also affect the stability of metapopulations. \citet{Hestbeck1988} suggested that the social fencing of immigrants by patch dwellers can stabilize dynamics by reducing dispersal-induced oscillations. Immigrants may also choose not to settle in an overcrowded patch because of increased resource competition; this behavior has been found to stabilize metapopulations \citep{RuxtonRohani1998}. Other behaviors such as conspecific attraction of immigrants and settlement facilitation have been argued to influence metapopulation dynamics as well \citep{Ray1991,Alvarado2001}.

The previous theoretical studies are based on models using specific functional forms. This restriction of the kinetics in the model is necessary to obtain certain results, such as steady state values of population densities. However, one may ask how the choice of a specific function affects the results. For example, \citet{Gross2004} showed that phenomena such as the paradox of enrichment can be strongly dependent on the particular functional form used in the model. \citet{Kisdi2010} also found that the effect of costly dispersal on stability was dependent on the growth function in use.

Other assumptions are frequently made in order to analyse the stability of metapopulations.
Because metapopulations lead to high-dimensional dynamical systems, previous mathematical studies typically reduced their complexity by assuming that dispersal is symmetric or patches are identical.
The influence of growth rate heterogeneity between patches has been investigated by \citet{Dey2006}, who found that it does not affect stability, even in different spatial topologies. According to this study, the effect of dispersal on the stability of metapopulations is thus not affected by the spatial arrangement of patches.

Here, we investigate the stability of metapopulations using a generalized modeling approach. 
We introduce a general metapopulation model, which does not assume specific kinetic laws, and encompasses both homogeneous and heterogeneous cases.
In the homogeneous case, all patches are identical (as in \citet{Ruxton1997a}), whereas in the heterogeneous case, demographic parameters may differ between patches \citep{Dey2006,StrevensBonsall2011}.
The homogeneous case enables us to draw analytical conclusions on the influence of dispersal behaviors such as density-independence, positive and negative density-dependence of dispersal, but also of less studied behaviors such as social fencing, settlement facilitation and conspecific attraction.
We also show that for heterogeneous webs the influence of density-independent dispersal on stability is not neutral, but strongly dependent on the topology of the metapopulation.

\section{Model}
\label{sec:1}

Metapopulations are classically approached using patch-occupancy mo\-dels, where local population size is ignored, and only the fraction of occupied patches is modeled \citep{Hanski1991}. However, deterministic within-patch dynamics models as the one we use here have proven particularly suited to study the influence of dispersal behaviors on metapopulation stability \citep{Taylor1990}.
We consider a metapopulation consisting of $M$ patches and denote the population density in patch $i$ by the scalar $X_i$.
The dynamics of the metapopulation can then be described by the following system of differential equations:

\begin{equation}
\label{model}
\dot{X}_{i}= G_{i}(X_i) - L_{i}(X_{i}) + \sum_{k=1}^{M} I_{i,k} (G_{i}(X_{i}),G_{k}(X_{k})),
\end{equation}

where $G$, $L$, and $I$ denote potentially non-linear functions governing the local growth, loss, and immigration rates in the respective patches. 
We distinguish between different immigration terms $I_{i,k}$ originating from different source patches $k \ne i$.

Note that in Eq. \ref {model} the immigration is assumed to depend on the growth rate of the donor patch and not on its density.
This choice is directly intuitive for a population with distinct life stages (see for instance \citet{Hassell1995}), where only a proportion of the juveniles migrates to other patches. 
Further, this formulation of the model enables to segregate competition and dispersal, in accordance with Hassell's criticism of Bascompte \& Sol\'{e} model \citep{BascompteSole1994,Hassell1995}.
Dispersal of juveniles is a common strategy in ecological populations, especially in animal species such as barnacles or hare \citep{Kent2003,Bray2007}.
In this model immigration is directly dependent on the number of juveniles in the donor patch. 
This enables to describe the density-dependence of dispersal, that is a widespread adaptive behavior in animal populations \citep{Travis1999}.

In the model, immigration is also assumed to depend on the growth rate in the recipient patch, as the number of juveniles of this patch can affect the success of immigration by inhibiting their settlement \citep{Hestbeck1988} or facilitating it \citep{Alvarado2001}. 
The choice of potential patches by immigrants is random as such, only the decision to settle or not depends on growth in recipient patches.
Note that when immigrants do not settle in a patch, they could potentially reach other patches and settle there. 
This would generate a feedback between growth in the potential recipient patch and immigration in other patches, which is not taken into account in our model. 

The equation above does not include a term for losses incurred by emigration. Depending on the ecological context emigration losses can be absorbed in either the loss or the gain term that are already included in the equation, by changing the interpretation of these terms accordingly. For instance, consider the scenario where a fixed proportion of juveniles/propagules $c$ leaves the patch to try to settle elsewhere. In this case the losses by emigration were $E= c G(x)$. We can include this loss directly in the growth function of the origin patch, such that we obtain a new growth function $\tilde{G}(x)=(1-c)G$. Thus emigration losses can be absorbed into the growth function by interpreting the $G$ that appears in the equation as ``growth after emigration''. 

When studying certain questions it is advantageous to absorb the emigration losses in the loss function instead. 
In particular, this formulation of the model facilitates the interpretation of the immigration function, that is dependent on the growth functions from the donor patches.
In this case we interpret the loss function as the sum of all losses, including emigration. 

In principle, one could also account for the net effect of emigration and immigration in a single dispersal function. However, this would necessarily lead to negative dispersal terms and interdependency between dispersal terms, both of which are incompatible with the mathematical procedure for stability analysis used below (see Appendix A). 

We remark that our model captures each local population only in a single variable and thus does not resolve the age or stage structure in the population. While we assume that juveniles or propagules play an important role in dispersal, we do not resolve their impact on local population dynamics. This choice is motivated by classical kinetic models of food webs where, say, the number of juveniles is not modelled explicitely as it becomes slaved to the number of adults and hence does not constitute an independent dynamical variable. To gain a general understanding of the impact of dispersal on stability, describing local populations by single variables can thus be justified. 
However, one should be aware that in specific scenarios, dynamics of juveniles may play an important role and has to be resolved explicitely by vector-valued variables \citep{Hastings1992,Castro2006}.

We note that Eq. \ref{model} remains applicable in a much wider class of settings. Because we do not restrict the functional forms of $G$, $L$, and $I$, letting the immigration rate depend on $G$ rather than $X$ does not reduce the generality as long as $G$ is a reversible function, which is quite generally the case. 

The generalized model from Eq. \ref{model} describes a whole class of conventional models, in which the functions $G$, $L$ and $I$ are restricted to specific functional forms. Many of these conventional models display steady states, i.e. states in which densities of populations remain locally stationary (see for instance \citet{Hassell1995}). However, not all of these steady states are necessarily stable to perturbations \citep{Kuehn2013}.
In the following we identify conditions that govern the stability for all positive steady states in all models within the class considered here. For this purpose, we proceed to a linear stability analysis of our model: the system is considered stable if it returns to its steady state after a small perturbation \citep{May1972,GrimmWissel1997}.

Let us emphasize that the approach used here assumes only that at least one feasible steady state exists somewhere in the space of models considered. We do not require that every model in this space has a feasible steady state, or that steady states are unique or stable. 
The analysis carried out in the following reveals the stability of all feasible steady states in the class of models. Following Gross \& Feudel \citeyearpar{GrossFeudel2006} and \citet{Yeakel2011} we denote the biomass densities in an arbitrary steady state by $(X_1^{*},X_2^{*},...,X_M^{*})$, and define the normalized variables 

 \begin{equation}
 \label{normvar}
x_i:=\frac{X_i}{X_i^{*}},
\end{equation}

and the normalized functions
\begin{equation}
\label{normfunc}
g_{i}(x_{i}):=\frac{ G_{i} (X_{i}^{*}  x_{i})}{G_{i}^{*}} 
\; \;
,
\; \;
l_{i}(x_{i}):=\frac{ L_{i} (X_{i}^{*}  x_{i})}{L_{i}^{*}} 
\;\;
,
\end{equation}

\begin{equation*}
\eta_{i,k}(x_i,x_{k}):=\frac{ I_{i,k} (G_{i} ( X_{i}^{*} x_{i}),G_k(X_{k}^{*} x_{k} ) )}{I_{i,k}^{*}} ,
\end{equation*}

where we use an asterisk ($*$) to denote the values that the functions assume in the steady state $X^*$. 
Normalized variables and functions are thus equal to $1$ at the steady state.
In other words the normalized variables and functions measure flows and densities in terms of multiples of their stationary value.
We can now rewrite the model in terms of the normalized variables and functions, by substituting the definitions Eqs. \eqref{normvar} and \eqref{normfunc} into Eq. \eqref{model}. We obtain

\begin{equation}
\label{normmodel}
\dot{x}_{i} = \frac{G_{i}^{*}}{X_{i}^{*}}  g_{i}(x_{i}) - \frac{L_{i}^{*}}{X_{i}^{*}}  l_{i}(x_{i})  +
\sum_{k=1}^{M} \frac{I_{i,k}^{*}}{X_{i}^{*}}  \eta_{i,k} (x_i,x_{k})
\end{equation}

The prefactors appearing in this equation are unknown constants and therefore can be interpreted as unknown parameters of the model. We define

\begin{equation*}
\alpha_{i}:= \frac{G_{i}^{*}}{X_{i}^{*}} + \sum_{k=1}^{M} \frac{I_{i,k}^{*}}{X_{i}^{*}}  = \frac{L_{i}^{*}}{X_{i}^{*}} 
\; \;
,
\; \;
\nu_{i}:= \frac{G_{i}^{*}}{G_{i}^{*} +  \sum_{k=1}^{M} I_{i,k}^{*}} = \frac{1}{\alpha_{i}} \frac{G_{i}^{*}}{X_{i}^{*}}
\end{equation*}

\begin{equation*}
\tilde{\nu_{i}}:=  1 - \nu_i = \frac{\sum_{k=1}^{M} I_{i,k}^{*}}{G_{i}^{*} +  \sum_{k=1}^{M} I_{i,k}^{*}} = \frac{1}{\alpha_{i}} \sum_{k=1}^{M}  \frac{ I_{i,k}^{*}}{X_{i}^{*}}
\; \;
,
\; \;
\theta_{i,j}:=\frac{I^{*}_{i,j}}{\sum_{k=1}^{M} I^{*}_{i,k} }  = \frac{1}{\alpha_{i} \tilde{\nu_{i}}}.\frac{I_{i,j}^{*}}{X_{i}^{*}}
\end{equation*}

\begin{table}
\centering
{\small
\caption{List of $GM$ parameters used in the model}
\begin{tabular}{llll}
Parameter & Interpretation & Simulation value & Simulation value  \\
		&			& (homogeneous case)	& (heterogeneous case) \\
\noalign{\smallskip}\hline\noalign{\smallskip}
Scale &  &&  \\
parameters&&& \\
\noalign{\smallskip}\hline\noalign{\smallskip}
$\alpha_i$ & Rate of biomass & $1$ & $1$ \\
& turnover in patch $i$&&\\
$\nu_i$ & Fraction of growth & $1-\tilde{\nu}_i$ & $1-\tilde{\nu}_i$\\
&due to production in $i$&&\\
$\tilde{\nu}_i$ & Fraction of growth & \lbrack0,1\rbrack & $\{ 0.3, 0.5, 0.7, 0.9  \}$\\
&due to immigration in $i$&&\\
$\theta_{i,j}$ & Relative weight of $j$ in the gain &$\frac{I_{i,j}}{\sum_{k} I_{i,k} } $ &  $\frac{I_{i,j}}{\sum_{k} I_{i,k} } $  \\
 & from immigration of $i$&& \\
\noalign{\smallskip}\hline\noalign{\smallskip}
Exponent && & \\
parameters&&&\\
\noalign{\smallskip}\hline\noalign{\smallskip}
$\phi_i$ & Sensitivity of primary production  &\lbrack0,2\rbrack& $\lbrack 0,2 \rbrack$ \\
&to density in $i$&&\\
$\mu_i$ & Sensitivity of loss  &\lbrack0,2\rbrack& 1 \\
& to density in $i$&&\\
$\omega_{i,j}$ & Sensitivity of immigration in $i$ &\lbrack0,2\rbrack& 1 \\
 & to growth in the donor patch $j$ &&   \\
$\zeta_{i,j}$ & Sensitivity of immigration from $j$ &\lbrack-1,1\rbrack& 0 \\
&  to growth in the recipient patch $i$ && \\
\noalign{\smallskip}\hline
\end{tabular}
\label{tab:param1}
}
\end{table}

These parameters all denote densities and biomass fluxes in the system at equilibrium and are easily interpretable in the context of the model; in accordance with Gross $\&$ Feudel \citeyearpar{GrossFeudel2006}, we call them \emph{scale parameters}.
The parameter $\alpha_i$ is the biomass turnover rate of individuals in patch $i$. As a characteristic timescale of the species, it is always positive.

The parameter $\nu_i$ represents the fraction of growth in patch $i$ resulting from production, whereas $\tilde{\nu}_{i}$ denotes the fraction of growth in patch $i$ resulting from immigration.
These parameters thus encode the impact of dispersal on growth in each patch at the steady state.
We can assume $\tilde{\nu}$ to be close to 0 in self-sustaining patches, whereas $\tilde{\nu}$ is close to 1 in patches where the population can only be maintained by strong immigration.
The fraction of growth due to immigration $\tilde{\nu}$ depends on the steady state densities in the different patches, and on the structure of the dispersal network. It can thus differ between patches, but for instance in metapopulation networks with equal number of links and equal densities in each patch, this parameter can also be identical for all patches.

The parameter $\theta_{i,j}$ denotes the relative weight of contribution of individuals from patch $j$ to the immigration to patch $i$.
For instance if a patch $i$ receives an equal density of immigrants from $n$ neighboring patches  then $\theta_{ij} = 1/n$ for the neighboring patches j, and  $\theta_{ij} =0$ for all other patches from which no immigrants are received.
The set of variables $\theta_{i,j}$ defines the network topology of the metapopulation.

Using the newly defined parameters we can write the generalized model as 

\begin{equation}
\label{genmodel}
\dot{x}_{i} =  \alpha_{i} \big[ \nu_{i}  g_{i}(x_{i}) -  l_{i}(x_{i})  +
 \tilde{\nu_{i}} \sum_{k=1}^{M} \theta_{i,k}  \eta_{i,k} (x_i,x_{k})  \big]
\end{equation}

To study the stability of the steady state, we calculate its Jacobian matrix at the equilibrium under consideration \citep{GuckenheimerHolmes1983}. The Jacobian matrix constitutes a local linearization of the system and contains derivatives of the normalized functions with respect to the state variables, evaluated at the equilibrium. Again these derivatives are unknown constants and can thus be considered as additional parameters of the system, which we call \emph{exponent parameters}. The exponent parameters are defined as 

\begin{equation*}
\phi_{i} :=  \left. \frac{ \partial{g_i(x_i)}  }{\partial{x_i}}  \right |_{x=x^*}
\; \;
,
\; \;
\mu_{i} :=\left. \frac{ \partial{l_i(x_i)}  }{\partial{x_i}}  \right |_{x=x^*}
\end{equation*}

\begin{equation*}
\zeta_{i,k} :=\left. \frac{ \partial{\eta_{i,k}(x_i,x_k)}  }{\partial{g_i}}  \right |_{x=x^*}
\; \;
,
\; \;
\omega_{i,k}: =\left. \frac{ \partial{\eta_{i,k}(x_i,x_k)}  }{\partial{g_k}}  \right |_{x=x^*}.
\end{equation*}

It can be shown that the exponent parameters are logarithmic derivatives of the original unnormalized laws and therefore measure the so-called elasticity of rate-laws in the steady state \citep{Yeakel2011}. Elasticities were originally proposed in economics as a measure of the nonlinearity of a function \citep{Nievergelt1983}. They have the advantage of allowing a particularly easy comparison with experimental data, and are now also widely used in metabolic control theory \citep{Fell1997}. 

Elasticities provide a nonlinear measure for the sensitivity of the functions to variations in the variables at the steady state.
For instance, power-law functions of the form $aX^p$ have a constant elasticity which is equal to $p$, e.g the elasticity of a linear function is $1$ and that of a quadratic function is $2$.
For more complex functions, the value of the elasticity is not constant, and depends on the steady state.
For instance for a Holling type-II functional response, the elasticity (with respect to prey) is 1 (linear) at low prey density and 0 (constant) close to saturation.

In the model, the parameter $\phi_i$ denotes the sensitivity of primary production in patch $i$ to the density of individuals. 
In populations that do not suffer from resource limitation, we would expect that production increases locally linearly with population size and thus that $\phi_i=1$. 
If social facilitation leads to an increase in productivity (a weak Allee effect \citep{TaylorHastings2005}), then even a superlinear increase in production with population size ($\phi_i>1$) can be observed.
But overcrowding can lead to a competition for resources, where the increase in production with population size becomes sublinear ($\phi_i<1$).
 The relationship between the computed elasticity of production to density and the functional forms of conventional models is discussed in \citep{Gross2004} and \citep{Yeakel2011}.

The parameter $\mu_i$ denotes the sensitivity of the loss rate to the density of individuals in patch $i$. 
In populations with resource limitation it can be expected that losses increase superlinearly with density ($\mu_i>1$).
On the other hand social facilitation can lead to a situation where per-capita loss rates decrease with increasing population density ($\mu_i<1$).

The parameters $\zeta_{i,k}$ and $\omega_{i,k}$ encode information on dispersal behaviors.
The parameter $\omega_{i,k}$ denotes the sensitivity of immigration to production in the donor patch.
Hence, when dispersal is density-independent, we expect that immigration increases linearly with production, and thus that $\omega_{i,k}=1$.
When patches are subject to resource competition, positive density-dependent emigration can be observed, where emigration is enhanced by overcrowding \citep{Matthysen2005}; in this case, we expect the sensitivity of immigration to growth in the donor patch to be superlinear ($\omega_{i,k}>1$). Positive density-dependent dispersal is well documented in several species, notably in butterflies \citep{NowickiVrabec2011} and mammals \citep{Matthysen2005}.
When however immigration is dampened by increased production in donor patches, dispersal can be described as negatively density-dependent. The sensitivity of immigration to growth in the donor patch is then sublinear ($\omega_{i,k}<1$).
This can be the result of a decreased emigration from donor patches. Such a behavior has been documented in butterfly species \citep{Baguette2010}.
Increased growth rates can also lead to resource depletion in donor patches, which can impede the success of immigration when dispersal is energetically costly \citep{Ruxton1997a}. The observed negative density-dependent dispersal is then the result of a costly dispersal, which is common particularly for small mammals such as hares \citep{GainesMcClenaghan1980}. 

The parameter $\zeta_{i,k}$ denotes the sensitivity of immigration to the growth rate in the recipient patch.
When growth inside the host patch has no influence on the settlement of immigrants \citep{LeGalliard2005}, settlement can be regarded as density-independent ($\zeta_{i,k}=0$).
When the settlement of immigrants is enhanced by high growth rates in the host patch, we expect that immigration increases with growth in the host patch ($\zeta_{i,k}>0$). This positive density-dependent settlement might be the result of settlement facilitation, as described in tunicate species \citep{Alvarado2001}, or of a conspecific attraction, observed in barnacles \citep{Kent2003}.
But when the settlement of immigrants is inhibited by a growing population in the recipient patch, immigration rate is a decreasing function of growth in the recipient patch ($\zeta_{i,k}<0$). This negative density-dependent settlement is similar to the social fence hypothesis, stating that when the neighbouring patches of a population grow in densities, they socially fence immigrants out by inhibiting their settlement \citep{Hestbeck1982}. This aggressive behavior is documented in microtine rodents \citep{Gundersen2002}. Negative density-dependent settlement might also originate from a decision not to settle in overcrowded patches \citep{RuxtonRohani1998}, a behavior that has been described in barnacles \citep{Kent2003}.

Elasticities enable the description of various dynamical behaviors, including dispersal behaviors. However, as some behaviors are not likely to co-occur, the value of elasticities is constrained by the plausibility of behavior combinations. In particular, some dispersal behaviors may have evolved in response to within-patch dynamics such as an increased competition for resources \citep{Travis1999,BowlerBenton2005}. It has thus been shown that positive density-dependent dispersal ($\omega>1$) can be associated to resource competition in the donor patch ($\phi<1$), while a negative density-dependent dispersal ($\omega<1$) can be associated to high growth in the donor patch ($\phi>1$) \citep{Kim2009}. The social fence hypothesis also suggests that an aggressive behavior against immigrants ($\zeta<0$) is associated to an increased competition for resources ($\phi<1$) \citep{Hestbeck1982}. 

Let us now illustrate how the values of the elasticities of dispersal relate to the functional forms of immigration that are found in conventional models. To do this, we focus on functional forms describing positive density-dependence of dispersal. \citet{Ruxton1996} describes immigration from patch $j$ to patch $i$ with a function of the form

\begin{equation}
I_{i,j}(G_j)=f(G_j)G_j, \:\:\: \mbox{with} \:\:\: f(G_j)=A(1-{\rm e}^{-\gamma G_j}),
\end{equation}

where $G_j$ is proportional to the number of juveniles in the donor patch, $f(G_j)$ is the fraction of juveniles that disperse, $A$ is the maximal fraction of emigrating juveniles, and $\gamma$ describes the strength of the density-dependence.
In order to compare this functional form to the elasticities of the generalized model, we compute the corresponding elasticity by following the procedure described in Gross \emph{et al.} (2004). This yields

\begin{equation}
\omega_{i,j}=1+ \frac{\gamma G_j^* {\rm e}^{-\gamma G_j^*}}{1-{\rm e}^{-\gamma G_j^*}}.
\end{equation}

Thus, $\omega_{i,j}$ is always greater than one, which indicates a positive density-dependent dispersal. In the limit where the saturation is small, $1-{\rm e}^{-\gamma G_j^*} \to 1$, a linear response, $\omega_{i,j} \to 1$, is recovered. In the opposite limit where the nonlinearity enhances emigration, $1-{\rm e}^{-\gamma G_j^*} \to 0$, the elasticity $\omega_{i,j}$ approaches infinity.

In the model of \citet{Silva2001}, immigration from patch $j$ to patch $i$ is described with a function of the form

\begin{equation}
I(G_j)=f(G_j)G_j, \:\:\: \mbox{with} \:\:\: f(G_j)=\frac{A G_j^k}{B^k+G_j^k},
\end{equation}

where $A$ is the maximum dispersal fraction, $B$ is the half-saturation value of dispersal fraction. The function $f$ is a Hill function which curve is described by the coefficient $k$.
The calculation of the elasticity of dispersal yields

\begin{equation}
\omega_{i,j}=1+ \frac{k}{1+\chi_j^k},
\end{equation}

where $\chi_j=G_j^{*k}/B^k$ is the growth rate measured in terms of multiples of the half saturation constant. The parameter $\omega_{i,j}$ is also always greater than one, indicating a positive density-dependent dispersal. 
We recover a linear relationship in the case $\chi _j\to \infty$, where the growth rate is much larger than the half-saturation constant. Conversely the strongest sensitivity $\omega_{i,j} = 1+k$ is found in the limit $\chi_j \to 0$, where the growth rate is very small.

These two examples highlight an advantage of generalized models in comparison with conventional models. Stability results of conventional models depend on the functional form of dispersal, and on the related steady states. On the other hand, scale parameters and elasticities of generalized models can describe an infinite number of steady states found with specific functional forms. 
We however emphasize that, as with any modeling approach, the utility of the generalized model is dependent on a realistic choice of parameter values. If parameter values are chosen from the realistic ranges identified here, it is guaranteed that the following is true: for every generalized parameter set, there is a family of realistic models within the class of models considered, that have a feasible steady state that is characterized by the given generalized parameters. As pointed out in \citep{Yeakel2011} at least one of these models can be constructed very easily. Thus for every generalized parameter set, we can immediately construct a model that is realistic and has a feasible steady state that is described by the parameter set. This is illustrated in Appendix B.

A summary of all parameters used here is shown in Table \ref{tab:param1}, and the ecological interpretation of the elasticities can be found in Table \ref{tab:param2}. The relationship between generalized models and conventional models is illustrated in \citep{Plitzko2012}. For additional discussions of generalized modeling, see \citep{GrossFeudel2006} and \citep{Gross2009}.

\begin{table}
\centering
{\small
\caption{Behavioral interpretation of the values of exponent parameters}
\begin{tabular}{ll}
Elasticity & Corresponding behavior  \\
value & \\
\noalign{\smallskip}\hline\noalign{\smallskip}
$\phi>1$ & positive density-dependent growth \\
$\phi <1$ & negative density-dependent growth\\
\noalign{\smallskip}\hline\noalign{\smallskip}
$\mu>1$ & positive density-dependent loss \\
$\mu <1 $ & negative density-dependent loss \\
\noalign{\smallskip}\hline\noalign{\smallskip}
$\omega=1$ & density-independent emigration \\
$\omega>1$ & positive density-dependent dispersal \\
$\omega<1$ & negative density-dependent dispersal \\
\noalign{\smallskip}\hline\noalign{\smallskip}
$\zeta=0$ & density-independent settlement \\
$\zeta>0$ & positive density-dependent settlement \\
$\zeta<0$ & negative density-dependent settlement \\
\noalign{\smallskip}\hline
\end{tabular}
\label{tab:param2}
}
\end{table}

We can now write the Jacobian matrix in an arbitrary steady state. 
The Jacobian matrix, $\rm \bf J$, with $J_{i,j}=\partial \dot{x_i} / \partial x_j$, captures the response of the focal steady state $X^*$ to sufficiently small perturbations.
The diagonal elements of the Jacobian matrix are 

\begin{equation}
\label{diagjac}
J_{i,i} = \alpha_i  \big[ \nu_i  \phi_i -\mu_i +   \tilde{\nu_{i}} \sum_{k=1}^{M} \theta_{i,k}  \phi_i  \zeta_{i,k}  \big],
\end{equation}

and the non-diagonal elements are 

\begin{equation}
\label{nondiagjac}
J_{i,j} = \alpha_i \big[ \tilde{\nu_i}  \theta_{i,j}  \phi_j \omega_{i,j} \big]
\end{equation}

A given steady state is stable if all eigenvalues of the Jacobian matrix have negative real parts.
We can thus evaluate the stability of the steady state under consideration by checking whether the largest eigenvalue has a positive real part \citep{GuckenheimerHolmes1983}.

\section{Homogeneous patches}
\label{sec:2}

We first consider the case of homogeneous metapopulations, where all patches are characterized by the same parameter values and are thus equivalent. 
Therefore, all row sums of the Jacobian matrix are equal to

\begin{equation}
\sum_{j=1}^{M} J_{i,j}= \alpha C, \:\:\: \mbox{with} \:\:\: C := \phi(1+\tilde{\nu}(\omega+\zeta-1))-\mu,
\end{equation}

because $\sum_{j=1}^{M} \theta_{i,j} =1$.
Matrices with identical row sums always have one eigenvalue that is identical to the row sum. For the matrices considered here this eigenvalue is indicative of stability (see Appendix). Thus, because $\alpha$ is positive, the system is stable when $C<0$ and unstable otherwise.

We notice that increasing the elasticity of loss ($\mu$) is always stabilizing. In absence of dispersal ($\tilde{\nu}=0$), increasing the elasticity of production ($\phi$) is always destabilizing. 
This confirms that a non-linear loss stabilizes the metapopulation, whereas a highly positive density-dependent production is destabilizing (cf \citet{Gross2009}).
In presence of dispersal ($\tilde{\nu}>0$), increasing $\phi$ can be stabilizing if $\omega + \zeta < (\tilde{\nu}-1)/\tilde{\nu}$.
For this inequality to hold, at least one of the elasticities of dispersal ($\omega$ or $\zeta$) must be negative. Intuitively, a nonlinear production can become stabilizing when it is dampened by dispersal.

To assess the influence of dispersal on stability, we compute the derivative of $C$ with respect to the fraction of growth due to immigration $\tilde{\nu}$. We obtain

\begin{equation}
\label{dernuhom}
\left. \frac{\partial{C}}{\partial{{\tilde{\nu}}}}  \right |_{*}  = \phi (\omega +\zeta -1)
\end{equation}

Because the sensitivity of production to density is generally positive ($\phi>0$), the sign of the derivative above depends only on the elasticities $\omega$ and $\zeta$.
In the case of density-independent dispersal ($\omega=1$ and $\zeta=0$) the derivative $\partial C/ \partial \tilde{\nu}$ vanishes such that dispersal has no effect on metapopulation stability.
A destabilizing effect is observed when $\omega + \zeta >1$. We thus expect to see a destabilizing effect of dispersal in systems with positive density-dependent dispersal ($\omega>1$) and positive density-dependent settlement ($\zeta>0$). This is intuitive as a strong positive feedback loop exists in this case. Conversely, dispersal has a stabilizing effect when $\omega + \zeta <1$. Which can be the case if dispersal ($\omega<1$) or settlement ($\zeta<0$) are negatively density-dependent. A summary of the impacts of dispersal behaviors on stability can be found in Table \ref{tab:discussion}.

\begin{figure}[!]
  \centering
  \subfloat[analytical result]{\includegraphics[height=0.35\textheight,width=0.48\textwidth]
  {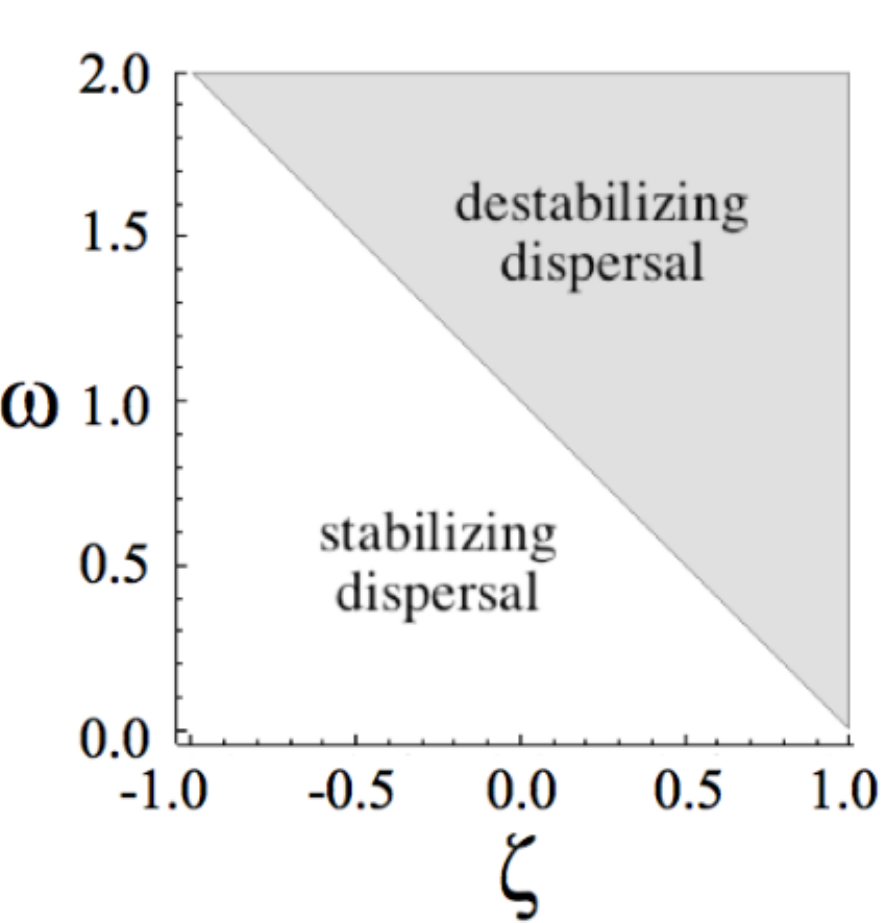}}  
   \subfloat[numerical result]{\includegraphics[height=0.35\textheight,width=0.54\textwidth]
  {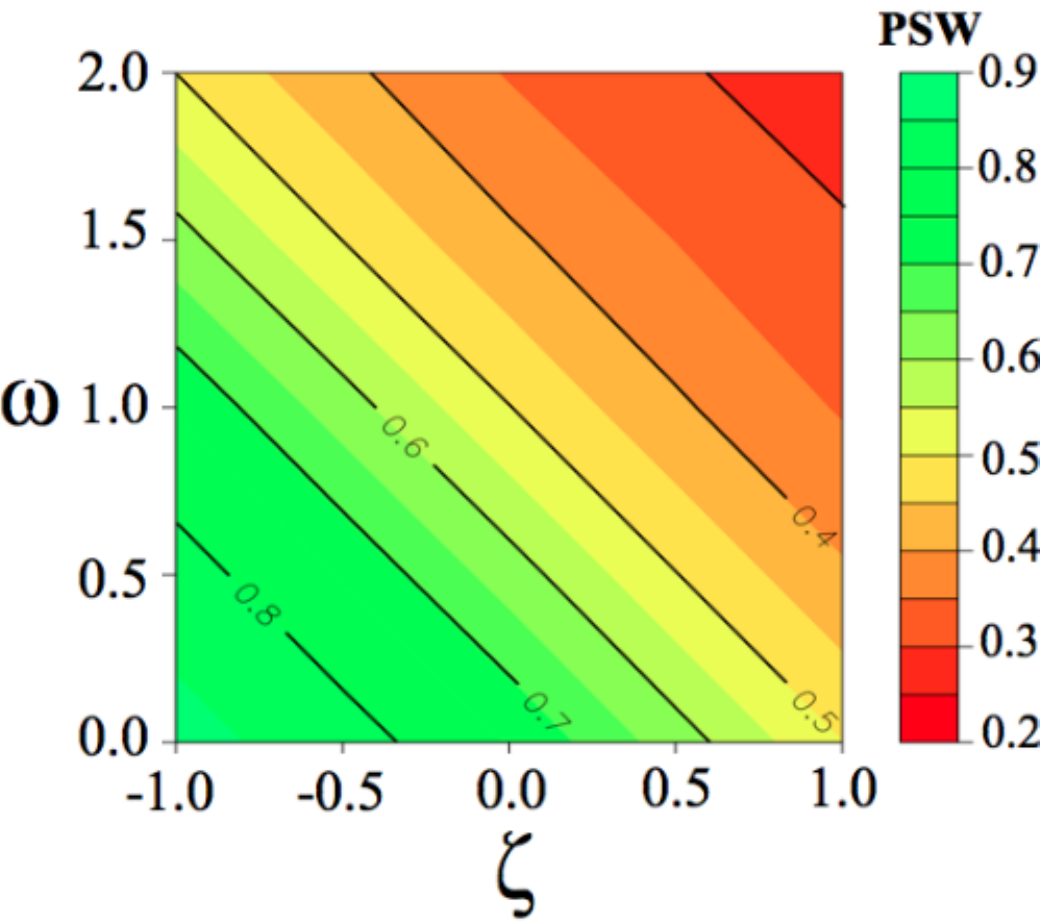}}  
 \caption{
  Influence of dispersal on stability for homogeneous patches.
Analytical results (a) show that dispersal is stabilizing for low values of both $\omega$, the elasticity of immigration with respect to the growth rate in the donor patch, and $\zeta$, the elasticity of immigration with respect to the growth rate in the recipient patch. 
Numerical results (b), based on the analysis of $10^4$ metapopulations show that the proportion of stable webs ($PSW$) decreases linearly as a function of parameters $\zeta$ and $\omega$. 
 \label{analysis}
 }
\end{figure}

We emphasize that both elasticities impact the nature of dispersal equally, such that a strong negative density-dependent settlement can possibly overcome a weak positive density dependence of dispersal, resulting in a stabilizing relationship. 
These results are also shown graphically in Fig. \ref{analysis}a.

Let us now investigate the effect of dispersal on homogeneous metapopulations numerically. For this purpose we first generate an ensemble of habitat topologies by placing $k$ dispersal links between $M$ habitat patches. The habitats thus form Erd\H{o}s-Renyi random graphs with connectance $C= k/(M(M-1))$. Disconnected habitats are rejected.
A randomly generated network can then be represented by its adjacency Matrix ${\rm \bf A}$, where $A_{ij}$ is equal to 1 if a link from patch i to j exists, and 0 otherwise. From the random adjacencies we compute the parameters 

\begin{equation}
\theta_{ij}=\frac{A_{ji}}{\sum_i A_{ji}},
\end{equation}

which is consistent with the assumption that $\theta_{i,j}$ denotes the relative weight of $j$ in the gain from immigration of $i$.
For each simulation, the parameters $\alpha$, $\omega$ and $\zeta$ are fixed, while the parameters $\tilde{\nu}$, $\phi$ and $\mu$ are randomly drawn from realistic ranges (See Table. \ref{tab:param1}). 
To investigate the stability of each network, we calculate the Jacobian matrix by using Eq. \ref{diagjac} and \ref{nondiagjac}.
For each pair of $\omega$ in the range $\lbrack 0,2 \rbrack$ and $\zeta$ in the range $\lbrack -1,1 \rbrack$, we generate a set of $10^4$ metapopulations, and we compute a proportion of webs ($PSW$) for which the generated metapopulations are stable (as in \citet{Gross2009}).
This measure describes the proportion of generalized models that are found to be stable for each pair of dispersal elasticities, provided that a feasible equilibrium exists for all realistic sets of parameters. As this measure changes with the elasticities of dispersal, it is an indicator of how these elasticities impact stability.

Results are shown in Fig.\ref{analysis}b, for metapopulations of 10 patches and a connectance of 0.5. 
The ranges of $\phi$ and $\mu$ were chosen such that in absence of dispersal ($\tilde{\nu}=0$) $50\%$ of the metapopulations are stable. Other ranges could potentially change the value of the $PSW$, but not the observed effect of the dispersal elasticities on it. When $\omega+\zeta=1$, we observe that dispersal has no effect on stability. Further, when $\omega+\zeta>1$, the proportion of stable webs is decreased by dispersal. Conversely, when $\omega+\zeta<1$, stability is increased by dispersal. These results are consistent with our analytical findings.

\section{Heterogeneous patches}
\label{sec:3}

Let us now investigate the heterogeneous case where parameter values may differ between patches.
To focus on the effects of heterogeneity, we consider particularly the case of density-independent dispersal and settlement ($\omega=1$, $\zeta=0$). As shown above, this type of behavior does not have any impact on the stability of homogeneous metapopulations. Any stabilizing or destabilizing effects observed below are therefore driven by heterogeneities in the population. 

In the following we study the effect of dispersal on heterogeneous metapopulations numerically.
We assume that heterogeneous metapopulations can emerge from differences in the conditions of growth between patches \citep{Dey2006}, driven by demographic heterogeneities.
In the example of a logistic growth, negative density-dependence ($\phi<1$) may arise close to the carrying capacity, while a positive density-dependent growth ($\phi>1$) is expected when densities are close to zero.
Thus, to create an heterogeneity between patches, we randomly draw $\phi_i$ from an uniform distribution of range $\lbrack 0,2\rbrack$, centered on 1.
The other elasticities are fixed, as is the fraction of growth due to immigration $\tilde{\nu}_i$ (see Table \ref{tab:param1}).
For each pair of $M$ and $C$ in the respective ranges, we generate a set of $10^4$ metapopulations, and we compute a proportion of webs ($PSW$) for which the generated metapopulations are stable.
The maximal number of patches is set to 15. In real metapopulations, this number can however be higher or lower, depending on the presence of suitable habitats and on dispersal distances \citep{Hanski1998}.

\begin{figure}[!]
  \centering
  \subfloat{\label{}\includegraphics[width=0.95\textwidth]
  {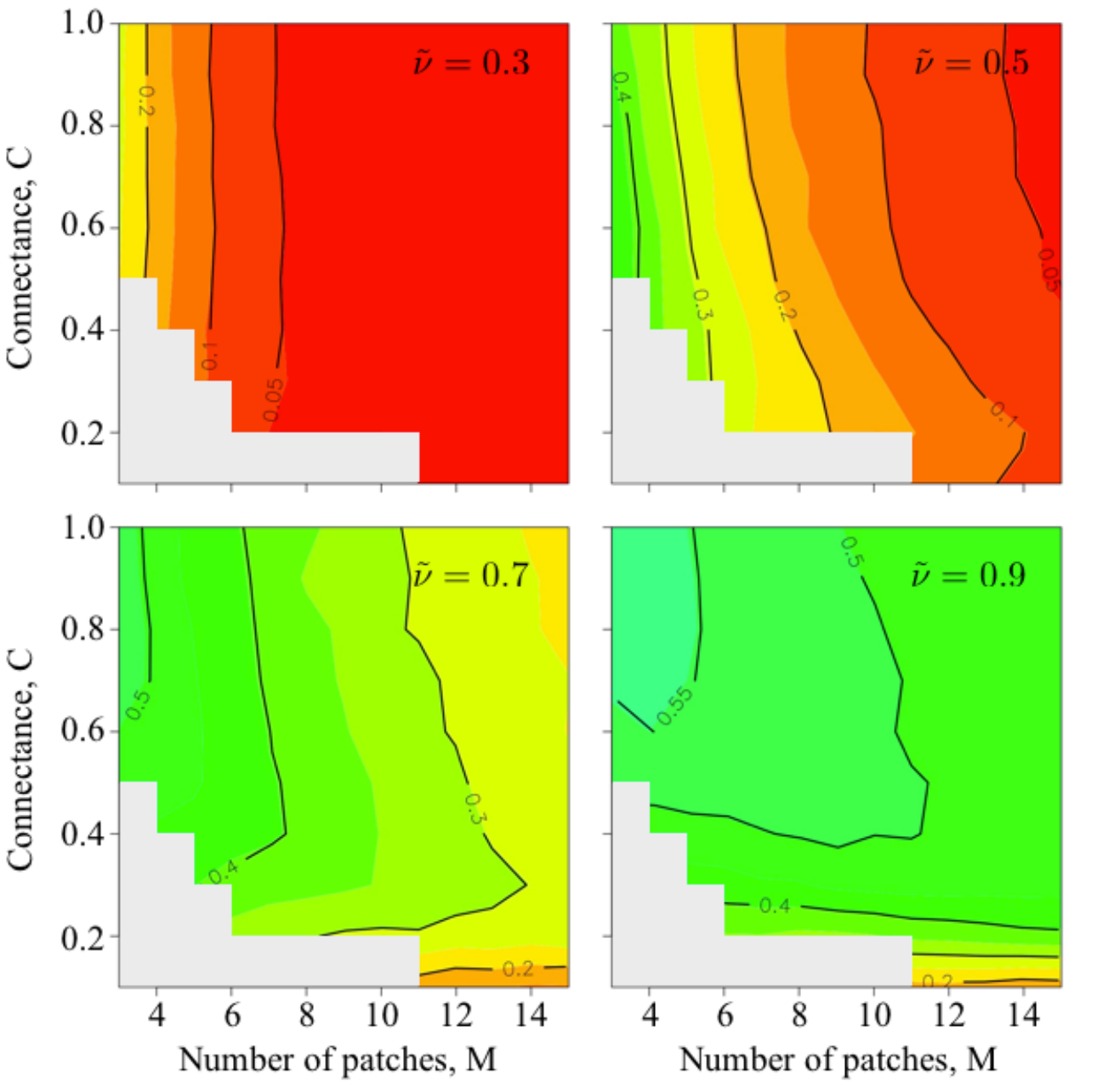}}  
 \caption{
  Proportion of stable webs ($PSW$) for heterogeneous metapopulations, in dependence on $M$ and $C$. Each panel corresponds to a different fraction of growth due to dispersal $\tilde{\nu}$. The grey area corresponds to infeasible networks. Dispersal is always stabilizing. The stabilizing effect depends on the interaction between the intensity of dispersal and the topology of the network. 
   \label{numerics_het}
 }
\end{figure}

Results are shown in Fig. \ref{numerics_het}.
Introducing dispersal between isolated heterogeneous populations is generally destabilizing.
The proportion of stable webs ($PSW$) of metapopulations is globally smaller than the $PSW$ of isolated metapopulations (excepted when $\tilde{\nu}=0.9$).
However, the correlation between the $PSW$ and the fraction of growth due to immigration is positive, indicating that the intensity of dispersal has a stabilizing influence on heterogeneous metapopulations.

Moreover, it is clear that the topology of metapopulations has an impact on their stability. Increasing the number of patches generally decreases the stability of metapopulations, so that a highly fragmented metapopulation is expected to be unstable. On the contrary, an increased connectance can increase the stability of metapopulations when the intensity of immigration is high. In that case, we expect small and highly connected metapopulations to be more stable than large and poorly connected ones.

We note that the impacts of the topology of heterogeneous metapopulations and the intensity of immigration are not independent. When dispersal is low ($\tilde{\nu}=0.3$), stability decreases with the number of patches, while connectance has almost no influence on stability.
When dispersal is high ($\tilde{\nu}=0.7$ and $\tilde{\nu}=0.9$), the number of patches has a greatly reduced impact on stability whereas connectance now has a stabilizing influence, especially when the number of patches is low. Note however that for high dispersal intensities, stability can also decrease with connectance when the number of patches is higher (e.g. when $M=10$).

We performed a multiple linear regression with the proportion of stable webs as the dependent variable. It confirms that the fraction of growth due to immigration has a significant stabilizing effect ($\beta_{\tilde{\nu}}=0.39\pm0.04$, $t(457)= 9.571$, $p<.001$), while the number of patches has a significant destabilizing effect ($\beta_{M}=-0.025\pm0.0025$, $t(457)= -10.223$, $p<.001$). The connectance is found to be generally destabilizing ($\beta_{C}=-0.11\pm0.036$, $t(457)=-3.137$, $p<.01$), in spite of its possible stabilizing effect when dispersal is high. In addition, it proves that while the interaction between $M$ and $C$ has no significant effect on stability ($t(457)=-1.374$, $p=0.17$), the interaction between $\tilde{\nu}$ and $M$ ($\beta_{\tilde{\nu}*M}=0.016\pm0.003$, $t(457)=5.34$, $p<.001$) and the interaction between $\tilde{\nu}$ and $C$ ($\beta_{\tilde{\nu}*C}=0.31\pm0.04$, $t(457)= 7.824 $, $p<.001$) have both significant stabilizing effects. These results show that the topology of the metapopulation can affect the impact of dispersal on stability. 
In particular, they suggest that the connectance and the number of patches reinforce the stabilizing effect of dispersal.

\begin{figure}[!]
  \centering
  \subfloat{\includegraphics[height=0.4\textheight,width=0.55\textwidth]
  {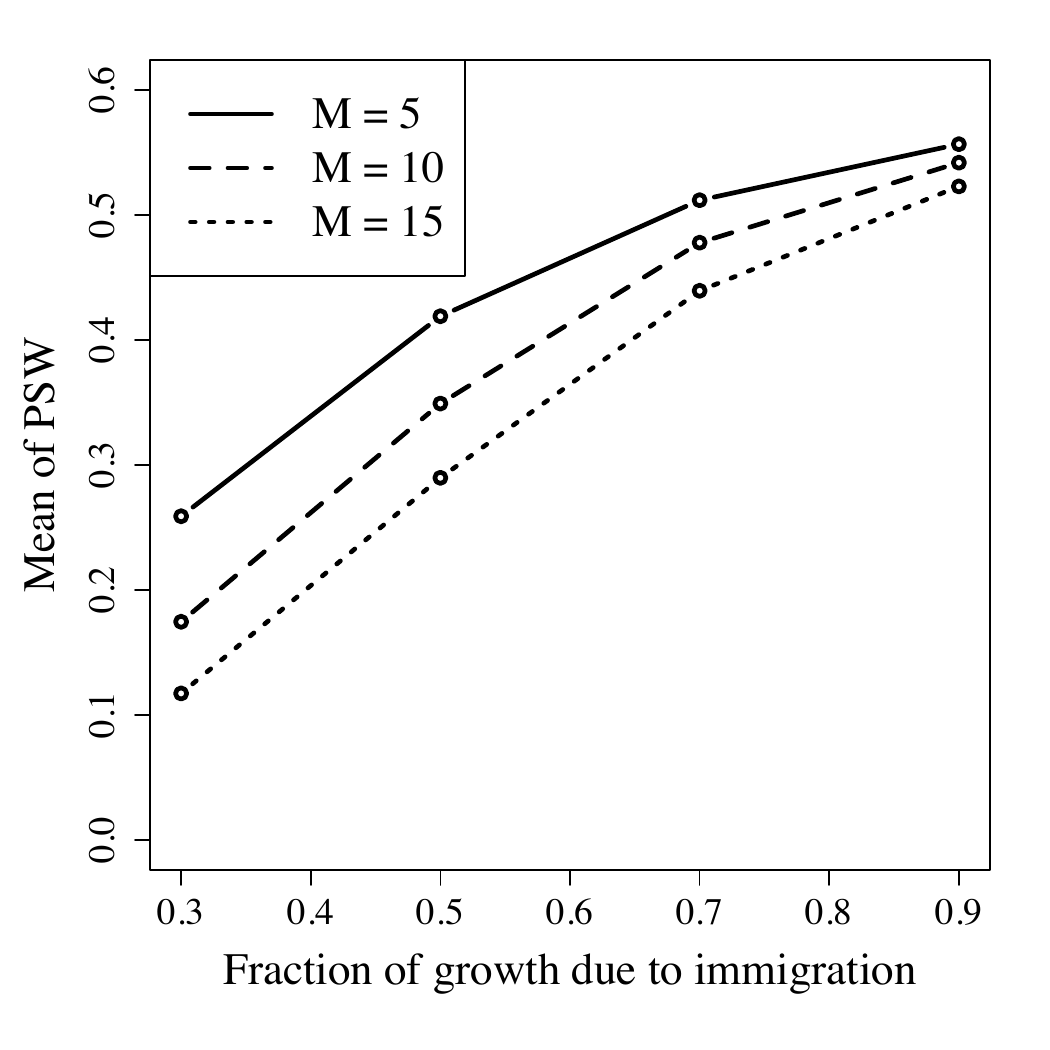}}  
   \subfloat{\includegraphics[height=0.4\textheight,width=0.55\textwidth]
  {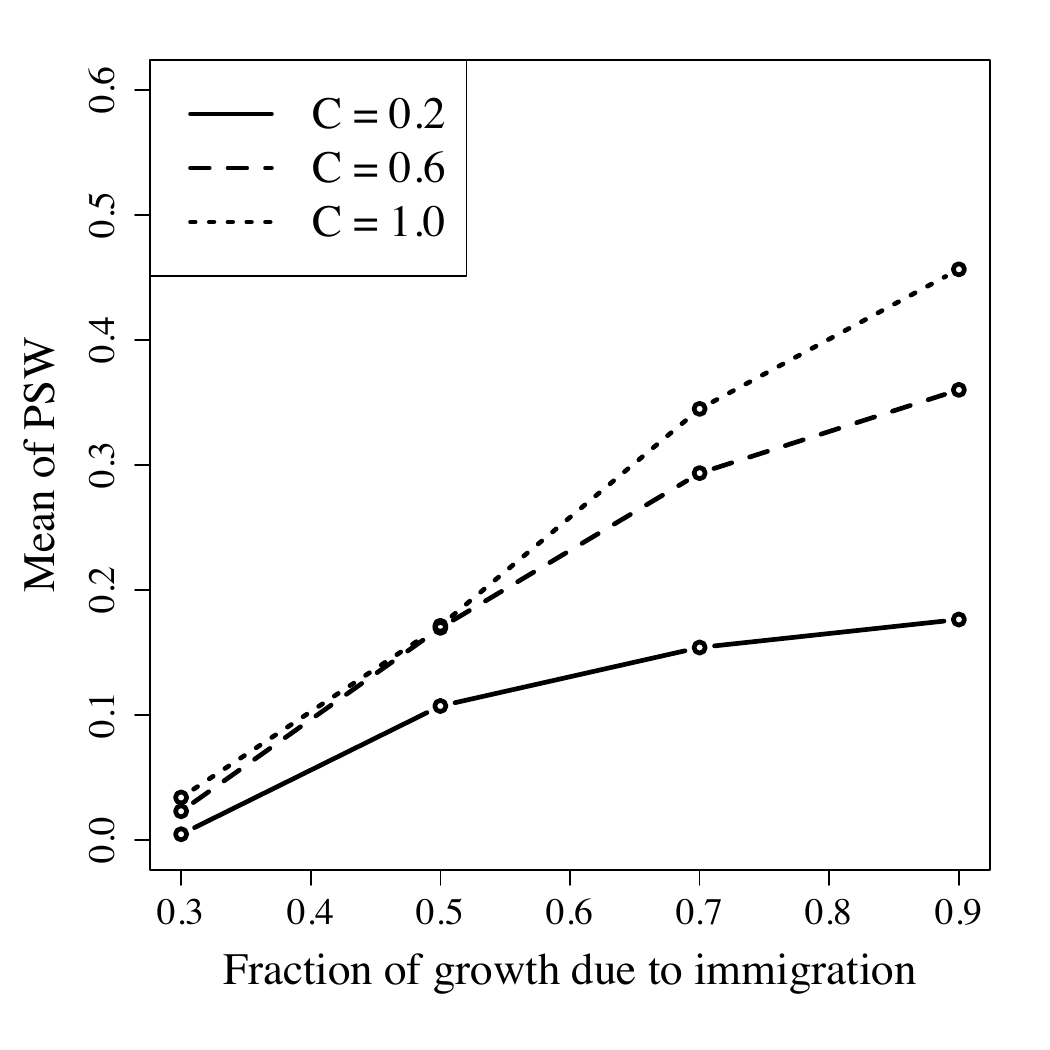}}  
 \caption{
  Interaction plots describing the interactive effects on stability of the fraction of growth due to immigration ($\tilde{\nu}$) and of network topology (number of patches $M$ and connectance $C$). Results are shown for 5, 10 and 15 patches and connectances of 0.2, 0.6 and 1. The left panel shows the interaction between the intensity of immigration and the number of patches. The right panel illustrates the interaction between the intensity of immigration and the connectance of the metapopulation network. When the intensity of immigration is high, the stabilizing effect of immigration increases with $M$ and $C$.
  \label{interacplot}
 }
\end{figure}

Two-way interaction plots show how precisely the stabilizing effect of immigration is affected by the topology of metapopulation networks (See Fig. \ref{interacplot}).
The slopes of the curves give information about the relative impacts of dispersal on the stability of the metapopulation. A positive slope that decreases with a topological parameter indicates that this parameter reduces the stabilizing effect of dispersal.
For a low intensity of immigration (low $\tilde{\nu}$), the number of patches and the connectance have no clear impact on the stabilizing effect of immigration. By contrast, for a high intensity of immigration (high $\tilde{\nu}$), the number of patches and the connectance increase the stabilizing effect of immigration. 
We thus observe that in heterogeneous metapopulations, the relationship between dispersal and stability crucially depends on the topology of metapopulations.

\section{Conclusions}
\label{sec:Discussion}

\begin{sidewaystable}
\caption{Comparison of the results with the existing litterature}
\label{tab:discussion}      
\begin{tabular}{llll}
Dispersal behavior & Effect of dispersal on stability & &  \\
\noalign{\smallskip}\hline\noalign{\smallskip}
 & Stabilizing & Destabilizing & No effect \\
\noalign{\smallskip}\hline\noalign{\smallskip}
Homogeneous metapopulation & & & \\
\noalign{\smallskip}\hline\noalign{\smallskip}
Density-independence &  & Bascompte \& Sol\'{e} \citeyearpar{BascompteSole1994} & Hassell \emph{et al.} \citeyearpar{Hassell1995}  \\
($\omega=1$, $\zeta=0$)&& \textbf{Our model when modified}&Rohani \citeyearpar{Rohani1996}\\
&& \textbf{(see Conclusions)} & Jang \& Mitra \citeyearpar{JangMitra2000} \\
&&& Jansen \& Lloyd \citeyearpar{JansenLloyd2000} \\
&&& \textbf{Our model} \\
&&&\\
Positive density-dependent && Ruxton \citeyearpar{Ruxton1996} & \\
dispersal ($\omega>1$, $\zeta=0$)& &Jang \& Mitra \citeyearpar{JangMitra2000}&\\
&&Silva \emph{et al.} \citeyearpar{Silva2001}&\\
&&Silva $\&$ Giordani \citeyearpar{SilvaGiordani2006}&\\
&&\textbf{Our model}&\\
&&& \\
Negative density-dependent &\textbf{Our model}& & \\
dispersal ($\omega<1$, $\zeta=0$) &&&\\
-\emph{negative density-dependent}&Ruxton \& Rohani \citeyearpar{RuxtonRohani1998}&&\\
\emph{emigration}&&&\\
-\emph{costly dispersal}&Ruxton \emph{et al.} \citeyearpar{Ruxton1997a}&Kisdi \citeyearpar{Kisdi2010}&\\
&Ruxton \emph{et al.} \citeyearpar{Ruxton1997b}&&\\
&&&\\
Positive density-dependent &&\textbf{Our model} & \\
settlement ($\omega=1$, $\zeta>0$)&&&\\
&&&\\
Negative density-dependent & Hestbeck \citeyearpar{Hestbeck1982} &  &    \\
 settlement ($\omega=1$, $\zeta<0$)& Hestbeck \citeyearpar{Hestbeck1988} &  &    \\
& \textbf{Our model} &  &\\
\noalign{\smallskip}\hline\noalign{\smallskip}
Heterogeneous metapopulation & & & \\
\noalign{\smallskip}\hline\noalign{\smallskip}
Density-independence& \textbf{Our model}  & & \\
($\omega=1$, $\zeta=0$)&(influence of the topology)&&  \\
\noalign{\smallskip}\hline
\end{tabular}
\end{sidewaystable}

We used generalized modeling to assess the influence of dispersal on the stability of metapopulations.
This approach allowed us to derive analytical results on a broad class of homogeneous networks and to obtain detailed results on heterogeneous metapopulations. 
Our analysis revealed six ecological results, that are gathered in Table \ref{tab:discussion}.

First, we found that, in accordance with most existing models, density-in\-de\-pen\-dent dispersal has no influence on the stability of homogeneous metapopulations.
If we follow Bascompte $\&$ Sol\'{e} assumption of not segregating reproduction and dispersal, i.e. if the immigration term depends on the density and not on the growth rate, then the conditions for a stabilizing dispersal depend on the sensitivities of immigration to density in the donor and recipient patches. Dispersal is then considered to be destabilizing when the sensitivity of production to density ($\phi$) is inferior to the sum of the elasticities of dispersal. Density-independent dispersal is thus destabilizing if growth is limited by competition ($\phi < 1$), which explains the results of Bascompte \& Sol\'{e} \citeyearpar{BascompteSole1994}.

Second, we showed that a positive density-dependent dispersal has a destabilizing effect on homogeneous metapopulations, which confirms previous findings on metapopulation stability \citep{Ruxton1996,JangMitra2000,Silva2001,SilvaGiordani2006}. 
Note that theoretical source-sink studies have also shown this behavior to promote the persistence of metapopulations \citep{Amarasekare2004}. This could help to understand why this behavior is widespread in birds and mammals \citep{Matthysen2005}.

Third, we found that a negative density-dependent dispersal has a stabilizing effect on homogeneous metapopulations. This can result from costly dispersal, that has already been suggested to stabilize metapopulations \citep{Ruxton1997a,Ruxton1997b}. Conversely, \citet{Kisdi2010} showed that costly dispersal can be destabilizing in homogeneous metapopulations, if the growth function of a single population is decreasing and sufficiently convex at the equilibrium. In the present model, this would imply the elasticity of growth $\phi$ to be negative; but in that case our calculation of the largest row sum would not be appropriate. A detailed exploration of the specific scenarios in which costly dispersal is destabilizing with the methodology proposed here is a promising target for future research.

Negative density-dependent dispersal can also result from a negative density-dependent emigration \citep{Baguette2010}, a behavior has been shown to influence the dynamics of metapopulations \citep{Amarasekare2004}. Our results suggest that negative density-dependent emigration might stabilize metapopulations. Further theoretical studies could help understand the precise effects of this behavior on metapopulation dynamics.

Fourth, we showed that a negative density-dependent settlement is stabilizing. This result is intuitive, as a negative density-dependent settlement stops crowded patches from being even more crowded and therefore tends to stabilize the metapopulation.
This result is consistent with Hestbeck's advocacy of a stabilizing effect of a social fence \citep{Hestbeck1982,Hestbeck1988}, who argues that when aggressiveness increases, immigration is inhibited, and oscillations due to dispersal are reduced.
It also supports the finding by \citet{RuxtonRohani1998} that immigrants' rejection of crowded patches is stabilizing.

Fifth, we suggested that a positive density-dependent settlement is destabilizing. This might result from settlement facilitation or conspecific attraction.
Evidence shows that settlement facilitation is widespread among marine organisms, and can occur through a modification of the environment \citep{Alvarado2001}, but this behavior remains poorly studied theoretically. Conspecific attraction has been extensively studied experimentally \citep{SerranoTella2003}, and has been shown to influence the dynamics of metapopulations \citep{Ray1991}. We emphasize that conspecific attraction must here be understood at the scale of the immigration phase of dispersal, and does not suppose that emigrants have information about the recipient patch.

Sixth, we showed that in heterogeneous metapopulations, density-in\-de\-pen\-dent dispersal can have a stabilizing effect.
We found that the importance of immigration in population growth and the topology of the metapopulation network conjointly modify the impact of dispersal on stability.
In particular, when the intensity of immigration is high, increasing the number of patches or the connectance reinforces the stabilizing effect of dispersal.
Contrary to \citet{Dey2006}, we thus argue that spatial heterogeneity of demographic parameters should not be neglected in modeling metapopulations (see for instance \citet{AbramsWilson2004}).

Results for heterogeneous webs follow from our definition of stability, that measures the ability of the system to return to the reference state after a temporary disturbance \citep{GrimmWissel1997}. This definition favours the survival of all local populations. Then, as increasing the number of patches while maintaining connectance equal augments the probability of local extinctions, stability can be expected to decrease with the number of patches. Likewise, as an increased immigration intensity might favor the maintenance of local populations, the fraction of growth due to immigration can be expected to have a positive effect of metapopulation stability. Other measures of stability such as persistence might however be expected to increase with the number of patches \citep{GyllenbergHanski1997}.

Though general, our model is not suited to represent phenomenons such as dispersal delays that can be significant in nature and have been shown to stabilize metapopulations \citep{Neubert2001,Klepac2007}. Time lags can also occur inside patches, where delays between settlement of juveniles and recruitment of new juveniles have been shown to destabilize open populations \citep{BenceNisbet1989}. Furthermore, we describe local dynamics as scalars, while local populations might have a more detailed structure, especially in age. The interaction between age structure and spatial structure has been shown to influence metapopulation stability, as dispersal rates may vary among age classes \citep{Hastings1992,Castro2006}. As \citet{Castro2006} pointed out, this effect is however dependent on conditions outside which our conclusions remain valid.

These results are based on a deterministic general model, although dispersal can be viewed as a highly stochastic process \citep{Abbott2011}. In particular, the dynamics of metapopulations are subject to both environmental and demographic stochasticity \citep{Hanski1991}. These two types of stochasticities create spatial heterogeneities that have been argued to favor the evolution of dispersal \citep{CohenLevin1991,Cadet2003}. In addition to making the analysis tractable, a deterministic specification is however relevant for large metapopulations where local fluctuations can be ignored.

In summary, we showed that dispersal can impact metapopulation stability through various dispersal behaviors, and that the effect of these behaviors on stability depends on the local within-patch dynamics of the metapopulation (as stressed in \citet{Amarasekare1998}).
We also showed that in more realistic heterogeneous metapopulations, already density-independent dispersal can affect stability.
We hope that the general analysis presented here can in the future form the basis of a more comprehensive classification of the effects of dispersal on metapopulation and metacommunity dynamics.

\section*{Acknowledgments}
This work has benefited from close collaboration with DFG Forschergruppe FOR1748.

\appendix
 \section{}
In this appendix we discuss the validity of replacing the leading eigenvalue with the row sum that we used in the analysis of homogeneous metapopulations. 

It is straightforward to confirm that all matrices with constant row sum $w$ have an eigenvalue $\lambda=w$ with a corresponding eigenvector (1,1,...). Furthermore the proof in \citep{Varga1962} shows that for non-negative irreducible matrices this eigenvalue is the largest eigenvalue of the matrix. This is an extension of the Perron-Frobenius theorem. The irreducibility condition is not an issue, as violating this condition would only introduce spurious zero eigenvalues that do not affect stability.

The non-negativity condition is slighty more serious as the matrices considered here generally contain negative elements.  However, according to Eq. \ref{diagjac} and \ref{nondiagjac} negative values can only appear in the diagonal elements. Let us assume the largest negative entry found is $z$. We now define a shifted matrix $J^z$ which is obtained by adding $z$ to each diagonal element. This has the effect of shifting the spectrum as a whole such that for every eigenvalue $\lambda$ of $J$ there is a corresponding eigenvalue $\lambda+z$ of $J^z$. 

Since the shifted matrix is an irreducible matrix with identical row sums $w+z$, we can apply the theorem from \citep{Varga1962} to show that the largest eigenvalue of this matrix is $w+z$. Because the spectrum of $J^z$ is identical to $z$ except for the constant shift this implies that the largest eigenvalue of $J$ is $w$.

This shows that for the class of matrices considered in the investigation of homogeneous food webs, the system is stable if the row sum is negative and unstable if the row sum is positive.  

\section{}
In this appendix we illustrate our approach by showing how a simple kinetic model with explicit dynamics can be drawn from a general model with realistic parameters.
Let us consider a 2-patch system and pick a random realistic parameter set, say

$v_1=0.7$
;
$\tilde{v_1} = 0.3$
;
$v_2 = 0.7$
;
$\tilde{v_2} =0.3$
;
$\phi_1=0.75$
;
$\phi_2=0.5$
;
$\alpha_1=1$
;
$\alpha_2=1$
;
$\omega_1=1$
;
$\omega_2=1$
;
$\zeta_1=0$
;
$\zeta_2=0$
;
$\mu_1=1$
;
$\mu_2=1$

This parameter set is not special and has been chosen to represent a fairly generic set with reasonable assumptions (e.g.~linear mortality). As $\phi_1$ and $\phi_2$ are not equal, the two patches are heterogeneous. We have chosen simple plausible values for the elasticities $\omega$, $\zeta$ and $\mu$ as these can fundamentally not impact our ability to create a system with the desired steady state.

From the parameters we know 

\begin{equation}
\begin{split}
\dot{x_1} = 0.7 g_1(x_1) + 0.3 \eta_{1,2}(x_2)  - l_1(x_1) \\
\dot{x_2} = 0.7 g_2(x_2) + 0.3 \eta_{2,1}(x_1) - l_2(x_2) 
\end{split}
\end{equation}

We can make the loss explicit by setting 

\begin{equation}
\begin{split}
\dot{x_1} = 0.7 g_1(x_1) + 0.3 \eta_{1,2}(x_2) - 0.6 e_{1,2}(x_1) - 0.4 m_1(x_1)\\
\dot{x_2} = 0.7 g_2(x_2) + 0.3 \eta_{2,1}(x_2) - 0.6 e_{2,1}(x_2) - 0.4 m_2(x_2)
\end{split}
\end{equation}

where $e_{i,j}$ is the normalized rate of emigration from patch $j$ to patch $i$ and $m_i$ is the normalized mortality rate in patch $i$. These normalized functions are equal to 1 at the equilibrium. In patch $1$ and $2$, the fraction of loss due to emigration is $0.6$. As we consider that emigration and immigration are linearly related to growth in the donor patches, we get:

\begin{equation}
\begin{split}
\dot{x_1} = 0.7 g_1(x_1) + 0.3 g_2(x_2)  - 0.6 g_1(x_1) - 0.4 m_1(x_1)\\
\dot{x_2} = 0.7 g_2(x_2) + 0.3 g_1(x_1) - 0.6 g_2(x_2) - 0.4 m_2(x_2)
\end{split}
\end{equation}

From these equations it appears that $50 \%$ of emigrants reach a patch. Now we have to find some functions that satisfy the remaining constraints. There are still very many different models that are consistent with the desired parameter values, so we can make the task of finding one easier for ourselves and pick a simple one. 

Suppose we wanted Holling-type-II-like gain rates

\begin{equation}
g_1(x_1) = A X_1/(K_1+X_1)
\end{equation}

For a simple example we chose functions that map 1 to 1 which can be done in this example by setting $A=(K_1+1)$ where we can still chose $K_1$ to create the desired value of $\phi$. We consider that mortality is a linear function of density. This procedure yields for instance

\begin{equation}
\begin{split}
\dot{x_1} = 0.7 (4x_1/(x_1+3)) + 0.3 (2x_2/(1+x_2))  - 0.6 (4x_1/(x_1+3)) - 0.4 x_1\\
\dot{x_2} = 0.7 (2x_2/(1+x_2)) + 0.3 (4x_1/(x_1+3)) - 0.6 (2x_2/(1+x_2)) - 0.4 x_2
\end{split}
\end{equation}

The advantage of choosing the functions to map 1 to 1 is that the normalization is the identity operation so that we can now straightforwardly replace the lower case letters with the unnormalized upper case letters

\begin{equation}
\begin{split}
\dot{X_1} = 0.7 (4X_1/(X_1+3)) + 0.3 (2X_2/(1+X_2))  - 0.6 (4X_1/(X_1+3)) - 0.4 X_1\\
\dot{X_2} = 0.7 (2X_2/(1+X_2)) + 0.3 (4X_1/(X_1+3)) - 0.6 (2X_2/(1+X_2)) - 0.4 X_2
\end{split}
\end{equation}

which is the desired model. This model assumes the following general form of metapopulation models

\begin{equation}
\dot{X_i} = G_i(X_i) + \sum_j \beta_{i,j} E_{j}(G_j(X_j))  - E_{i}(G_i(X_i)) - M_i(X_i),
\end{equation}

where $E_{i}$ describes emigration from patch $i$ and $\beta_{i,j}$ is the fraction of emigrants from patch $j$ that reach patch $i$. As dissipation is generally supposed to occur at a base level, $\sum_i \beta_{i,j}$ is inferior to 1. With mortality during dispersal, this $\beta$ could be negatively dependent on growth in the donor patch. 

Say for example that $40\%$ of emigrants (for instance coral larvae) dissipate at a base level. Then, an increased growth in the donor patch will increase the number of losses, but the proportion of lost emigrants will remain the same. With dispersal mortality due to costly dispersal, the proportion of lost emigrants will rise, thus affecting the dynamics of the system. On the contrary, behaviors such as conspecific attraction are expected to decrease the proportion of lost emigrants when growth increases.

Likewise we could construct such a model for every given plausible set of generalized parameters. When parameters are chosen such that immigration depends nonlinearly on productivity, then these nonlinearities can always be accommodated in the rate of settling success, which is ecologically reasonable. 

\section*{References}
  \bibliographystyle{elsarticle-harv} 
  \bibliography{tromeur_rudolf_gross_jtb.bib}

\end{document}